\documentclass{desyproc}

\begin{document}
\title{\vspace{-3cm}
\hfill{\small{DESY 08-149}}\\[2cm]
From Axions to Other WISPs}

\author{{\slshape Andreas Ringwald}\\[1ex]
DESY, Notketra{\ss}e 85, 22607 Hamburg, Germany}

\contribID{ringwald\_andreas}

\desyproc{DESY-PROC-2008-02}
\acronym{Patras 2008} 
\doi  

\maketitle

\begin{abstract}
We illustrate, taking a top-down point of view, how axions and other
very weakly interacting sub-eV particles (WISPs) arise in the course of 
compactification of the extra spatial dimensions in string/M-theory. 
\end{abstract}


It is a tantalizing question whether there is new physics {\em below}  
the Standard Model. That is to ask, whether there are new very 
light particles -- apart from the known ones with sub-eV mass, the photon and the neutrinos -- 
which are very weakly coupled to the Standard Model. 
In fact, embeddings of the latter into more unified theories, 
in particular into string theory, suggest their possible existence 
in a so-called hidden sector of the theory. Prominent examples of inhabitants of
the latter are the axion and axion-like particles, arising as pseudo Nambu-Goldstone
bosons associated with the breakdown of global anomalous
U(1) symmetries. They occur generically in realistic string compactifications, as we
will review below. Extra, hidden U(1) gauge bosons are also frequently encountered
in string embeddings of the Standard Model, as we will summarize below. 
There is no reason why some of these hidden U(1)
gauge bosons can not be massless or very light, in which case they also belong to 
the class of very weakly interacting sub-eV particles (WISPs). Further candidates for
WISPs are very light hidden sector particles which are charged under the hidden U(1)s. 

In this contribution, we will take a top-down point of view:
we will illustrate how axions and other WISPs arise in the course of compactification
of the extra dimensions of string theory. For the bottom-up point of view, i.e. for 
arguments and phenomenological as well as cosmological hints which point to the 
possible existence of WISPs, see the contributions of Joerg Jaeckel and 
Javier Redondo in these proceedings.  


{\em Axions from string compactifications.--}
The low-energy effective actions describing the dynamics of the massless
bosonic excitations of the heterotic and type II string theories in 9+1 dimensions 
are summarized in Table~\ref{tab:sugra}. 
\begin{table}
\centerline{\begin{tabular}{|l|l|}
\hline
Heterotic  & 
$S_{\rm H} 
=\frac{2\pi M_s^8}{g_s^2}\int d^{10}x \sqrt{-g}R 
-\frac{M_s^6}{2\pi g_s^2}\int\frac{1}{4}{\rm tr} F\wedge\star F
-{\frac{2\pi M_s^4}{g_s^2}\int\frac{1}{2} H\wedge\star H}+\ldots$
\\
\hline
Type II & 
$S_{\rm II} 
=\frac{2\pi M_s^8}{g_s^2}\int d^{10}x \sqrt{-g}R 
-\frac{2\pi M_s^{p+1}}{g_s}\int d^{p+1}x {\rm tr}\sqrt{-\det\left({g+B+F/(2\pi M_s)}\right)}$
\\ 
& 
\mbox{} \hspace{12ex}
$-2\pi {\rm i} M_s^{p+1} \int_{{\rm D}p} 
{\rm tr}\exp\left({B+F/(2\pi M_s)}\right)\wedge \sum_q C_q+\ldots$
\\
\hline
\end{tabular}}
\caption{Low-energy effective actions describing the dynamics of the massless bosonic 
excitations in the weakly coupled heterotic (top) and type II string theories with  
D$p$-branes in 9+1 dimensions. 
$R$ is the Ricci scalar, $F$ is the field strength of the gauge fields, and $H$ is the field 
strength of the two-form field $B$. In our conventions, $M_s=1/(2\pi\sqrt{\alpha^\prime})$,
with string tension $\alpha^\prime$. 
}
\label{tab:sugra}
\end{table}
As we will see, after compactification of six of the spatial dimensions, 
pseudo-scalar fields $a$ will generically arise which have a 
coupling $a\, {\rm tr}\,G \wedge G$ to the gluon field strength $G$ in the effective 
Lagrangian describing the low-energy dynamics of the theory in 3+1 dimension
and possess an anomalous Peccei-Quinn global shift symmetry, $a\to a+\epsilon$. 

These are the properties needed for the axionic solution of the strong CP 
problem~\cite{Peccei:1977hh}.   
Indeed, the anomalous shift symmetry implies that the axion field can enter in 
the low-energy Lagrangian only through derivative and explicit symmetry violating 
terms originating from chiral anomalies, 
\begin{eqnarray}
{\mathcal L}_a = 
\frac{1}{2} \partial_\mu a \partial^\mu a 
+ {\mathcal L}_a^{\rm int} \left[\frac{\partial_\mu a}{f_a};\psi\right]  
+ \frac{r\alpha_s}{4\pi f_a}\, a\, {\rm tr}\, G^{\mu\nu} {\tilde G}_{\mu\nu}+  
\frac{s\alpha}{8\pi f_a}\, a\,F^{\mu\nu} {\tilde F}_{\mu\nu}+\ldots \,, 
\label{axion_leff}
\end{eqnarray}
with dimensionless constants $r\neq 0$ and $s$, the (conventionally normalized) 
electromagnetic (gluonic) field strength $F$ ($G$), 
and the axion decay constant $f_a$. The CP violating term 
$
\alpha_s/(4\pi )\, \bar\theta\, {\rm tr}\, G_{\mu\nu} {\tilde G}^{\mu\nu} 
$
in the QCD Lagrangian can then be eliminated by exploiting the shift symmetry, 
$a\to a - \bar\theta f_a/r$: the $\bar\theta$ dependence is wiped out by the axion, 
providing a natural explanation why e.g. the electric dipole moment of the 
neutron is so small. 
Finally, the topological charge density $\propto \langle {\rm tr}\, G^{\mu\nu} {\tilde G}_{\mu\nu} \rangle \neq 0$,
induced by topological fluctuations of the gluon fields such as QCD instantons,  
provides a nontrivial potential for the axion field, giving a small mass
to the axion~\cite{Weinberg:1977ma}, which can be inferred via current algebra and expressed in
terms of the light ($u,d$) quark masses, the pion mass $m_\pi$ and the pion decay
constant $f_\pi$,   
\begin{eqnarray}
m_a =
         \sqrt{m_u m_d}/(m_u+m_d)\, 
m_\pi f_\pi/(f_a/r)
\simeq { 0.6\,  {\rm meV}} 
         \times
         \left( 
         10^{10}\, {\rm GeV}/(f_a/r)\right) \,.  
\end{eqnarray} 
For large axion decay constant $f_a$, we see that the axion is a prime
example for a WISP: it is very weakly interacting (cf. Eq.~\eqref{axion_leff}) 
and it is very light~\cite{Kim:1979if}. 
For various astrophysical, cosmological, and laboratory limits on $f_a$ arising from the
couplings of the axion to the Standard Model particles according to Eq.~\eqref{axion_leff}, 
see other contributions in these proceedings. Typically, for axions,
the limit is $f_a/r\gtrsim 10^{9}$~GeV.
%
Here, we will turn now to predictions of $f_a$ in string embeddings of the Standard Model.

In the compactification of the weakly coupled heterotic string, a universal, {\em model-independent}
axion appears as the dual of the antisymmetric tensor field $B_{\mu\nu}$ 
(whose field strength has been denoted by $H$ in Table~\ref{tab:sugra}), 
$da\sim \star dB_{\mu\nu}$, 
with $\mu$ and $\nu$ tangent to 3+1 dimensional Minkowski space-time~\cite{Witten:1984dg}.
Its decay constant $f_a$ is quite independent of the details of the compactification. In fact,
after compactification of the theory, originally described in 9+1 dimensions by  
$S_{\rm H}$ in Table~\ref{tab:sugra},  
on a 6 dimensional manfiold with volume $V_6$, the
resulting effective action can be matched to its standard normalization in 3+1 dimensions, 
\begin{eqnarray}
S_{\rm 3+1}
=\frac{M_P^2}{2}\int d^4x\sqrt{-g}\,R 
-\frac{1}{4g_{\rm YM}^2}\int d^4x\sqrt{-g}\,{\rm tr}\, F_{\mu\nu} F^{\mu\nu}
-{\frac{1}{f_a^2}\int\frac{1}{2} H\wedge\star H} + \ldots\,, 
\end{eqnarray}
with
\begin{eqnarray}
M_P^2 = (4\pi/g_s^2) M_s^8 V_6; \hspace{2ex}
g_{\rm YM}^2 = 4\pi g_s^2/(M_s^{6} V_6); \hspace{2ex}
{ f_a^2=g_s^2/(2\pi M_s^4  V_6)}\,,
\label{heterotic_couplings}
\end{eqnarray}
expressing the reduced Planck mass $M_P=2.4\times 10^{18}$~GeV, 
the gauge coupling $g_{\rm YM}$, and the axion decay 
constant $f_a$ in terms of the string coupling $g_s$, the string scale $M_s$, and the volume
$V_6$. Eliminating the volume $V_6$ and the string scale by means of the first two relations 
in Eq.~\ref{heterotic_couplings}, 
we end up with an axion decay constant of order 
of the GUT scale~\cite{Choi:1985je}, 
\begin{eqnarray}
f_a/r=\alpha_{\rm YM} M_P/(2\pi \sqrt{2})
\simeq  1.1\times 10^{16}\ {\rm GeV}\,,
{\rm \ for\ } \alpha_{\rm YM}=g^2_{\rm YM}/(4\pi)\sim 1/25\,. 
\label{f_a_heterotic}
\end{eqnarray}

{\em Model-dependent} axions arise in the context of weakly coupled heterotic strings  
from massless excitations of the two-form $B$-field on the 6 dimensional compact manifold~\cite{Witten:1984dg}. 
Correspondingly, their 
properties depend much more on the details of the compactification. Nevertheless, 
a recent exhaustive study has elucidated~\cite{Svrcek:2006yi} that also in this case
the axion decay constant cannot be smaller than $10^{15}$~GeV. 
Similar conclusions have been drawn for the 
axions in strongly coupled heterotic string theory~\cite{Svrcek:2006yi}. 
These findings can be easily understood physically: it is the string scale $M_s$ which mainly 
determines the axion decay constant~\cite{Conlon:2006tq}. And in the heterotic case, 
this scale is large, e.g. $M_s = \sqrt{\alpha_{\rm YM}/(4\pi)} M_P$ for the weakly coupled
heterotic string (cf. Eq.~\eqref{heterotic_couplings}). 

\begin{figure}[hb]
\centerline{\includegraphics[width=0.31\textwidth]{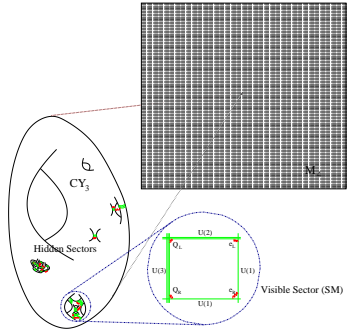}}
\caption{In compactifications of type II string theories the Standard Model
is locally realized by a stack of D-branes wrapping cycles in the compact dimensions. 
In general, there are also hidden sectors localized at different places. 
Light visible and hidden matter particles arise from strings located at intersection loci and stretching between brane stacks.   
Adapted from Ref.~\cite{Conlon:2006wz}.}\label{Fig:type_ii_comp}
\end{figure}

This may be different in compactifications of type II string theories which give rise to 
``intersecting brane worlds". In these theories, the Standard Model lives on a stack
of D$(3+q)$-branes which are extended along the 3+1 non-compact dimensions and wrap 
$q$-cycles in the compactification manifold (see Fig.~\ref{Fig:type_ii_comp}), 
while gravity propagates in the bulk, leading to a possibly smaller string scale at the 
expense of a larger compactification volume, $M_s\sim g_s M_P/\sqrt{V_6 M_s^6}$.   
In type II string theory, the axions come from the massless excitations of the
$q$-form gauge field $C_q$ (cf. Table~\ref{tab:sugra}). The precise predictions depend
on the particular embedding of the Standard Model~\cite{Conlon:2006tq,Svrcek:2006yi}, but generically one finds that the axion decay constant, $f_a\sim M_s$, can be substantially 
lower than in the heterotic case and in a phenomenologically very interesting range, e.g.  
\begin{equation}
f_a\sim M_P/\sqrt{V_6 M_s^6}\sim 10^{11}\ {\rm GeV},\  
{\rm for\ } 
V_6 M_s^6 \sim 10^{14}\,,
\end{equation} 
in LARGE volume flux compactification models~\cite{Conlon:2006tq}.


{\em Other WISPs: Hidden U(1)s and hidden matter.--} 
Additional hidden sector U(1) gauge factors are a generic feature of string compactifications. 
For example, in the ``mini-landscape" of orbifold compactifications of the heterotic string~\cite{Lebedev:2006kn} one encounters, at the compactification scale, a breaking of the gauge symmetry to 
a theory involving many hidden U(1)s, e.g.  
${\rm E}_8\times {\rm E}_8\to {\rm G}_{\rm SM}  \times {\rm U(1)}^4 \times 
[{\rm SO(8)}\times {\rm SU(2)}\times {\rm U(1)}^3]$ and the like. Similarly, as illustrated in
Fig.~\ref{Fig:type_ii_comp}, the type II compactifications generically invoke hidden sector
U(1)s\footnote{Not shown are possible U(1)s 
arising from branes wrapping bulk cycles and intersecting the SM branes. 
For a large volume of the bulk, these interact very weakly with 
the SM~\cite{Burgess:2008ri} and are thus WISP candidates.}, often also for global consistency requirements. 
Some of these hidden U(1)s may remain unbroken down to very small scales~\cite{Abel:2006qt}. 
In this case their dominant interaction with 
the Standard Model will be through kinetic mixing with the hypercharge U(1)$_Y$, 
described by the term
\begin{equation}
{\mathcal L} \supset \frac{\chi}{2 g g^\prime} {\hat Y}_{\mu\nu} {\hat X}^{\mu\nu}\,,
\end{equation}
in the low energy effective Lagrangian, where $Y_{\mu\nu}$ 
($X_{\mu\nu}$) is the hypercharge (hidden) U(1) 
field strength  and $g$ ($g^\prime$) is the hyper- (hidden-) charge.  
Often there 
is also light hidden matter charged under the hidden U(1)s, as illustrated for the type II
compactifications in Fig.~\ref{Fig:type_ii_comp}. After diagonalization of the gauge kinetic
terms by a shift $\hat X\to \hat X + \chi \hat Y$ and a multiplicative hypercharge renormalization, 
one observes that the hidden sector particles acquire a minihypercharge 
$g_h = \chi g^\prime$~\cite{Holdom:1985ag}. There are strong astrophysical limits, 
$g_h \lesssim 10^{-14}$, for masses below a few keV, as reviewed by Javier Redondo in 
these proceedings, and there are a number of ideas to probe such values in the laboratory 
as summarized by Joerg Jaeckel. Here, we would like to concentrate on the 
string theory predictions for $\chi$, which turn out to be comfortably small, but still
of phenomenological interest.

Kinetic mixing is generated by the exchange of heavy messengers that couple both
to the hypercharge U(1) as well as to the hidden U(1). In the context of compactifications 
of the heterotic string, its size has been estimated as~\cite{Dienes:1996zr}
\begin{equation}
\chi \sim g g^\prime /(16\pi^2)\,C\,\Delta m/M_P\gtrsim 10^{-17}\,,
{\rm \ for\ } C\gtrsim 10,\ \Delta m\gtrsim 100\ {\rm TeV}\,,
\end{equation}
where $\Delta m$ is the mass splitting in the messenger sector. 
Small values for $\chi$ can also be accommodated in type II compactifications. Here, kinetic
mixing can be understood as originating from the exchange of closed strings through the 
bulk~\cite{Lust:2003ky}. Correspondingly, it can experience a volume 
suppression~\cite{Abel:2006qt}, e.g., from D3-brane mixing, 
\begin{equation}
\chi \sim  g g^\prime/(16\pi^2)\,(V_6 M_s^6)^{-2/3}
\sim 10^{-14}\,,  
{\rm \ for\ } V_6 M_s^6\sim 10^{14} \,.
\end{equation} 
Exponentially suppressed values can be naturally obtained in flux compactifications with warped throats~\cite{Abel:2006qt}. Intriguingly, values even as small as  $\chi\sim 10^{-25}$
may be of phenomenological interest in the context of decaying dark matter~\cite{Chen:2008yi}.




\begin{footnotesize}


\end{footnotesize}


\end{document}